\documentclass{article}
\usepackage[utf8]{inputenc}

\usepackage{amsmath}
\usepackage{amssymb}
\usepackage{bm}
\usepackage{textcomp}
\usepackage{verbatim}

\usepackage{graphicx}
\usepackage{subfig}
\usepackage{float}
\usepackage{wrapfig,lipsum,booktabs}

\usepackage[colorlinks,citecolor=blue,urlcolor=blue]{hyperref}

\usepackage{fullpage}

\usepackage{natbib}

\title{A case study of glucose levels during sleep using fast function on scalar regression inference}

\author{Renat Sergazinov$^{1}$, Andrew Leroux$^{2}$, Erjia Cui$^{3}$, Ciprian Crainiceanu$^{3}$,\\
R. Nisha Aurora$^{4}$, Naresh M. Punjabi$^{5}$, Irina Gaynanova$^{1,*}$\\
\\
$^{1}$Department of Statistics, Texas A\&M University\\
$^{2}$Department of Biostatistics \& Informatics, University of Colorado Anschutz Medical Campus\\
$^{3}$Department of Biostatistics, Johns Hopkins University\\
$^{4}$ Rutgers Robert Wood Johnson Medical School\\
$^{5}$ University of Miami, Miller School of Medicine\\
$*$: irinag@stat.tamu.edu}

\date{}

\begin{document}

\maketitle

\begin{abstract}
Continuous glucose monitors (CGMs) are increasingly used to measure blood glucose levels and provide information about the treatment and management of diabetes. Our motivating study contains CGM data during sleep for 174 study participants with type II diabetes mellitus measured at a 5-minute frequency for an average of 10 nights. We aim to quantify the effects of diabetes medications and sleep apnea severity on glucose levels. Statistically, this is an inference question about the association between scalar covariates and functional responses. However, many characteristics of the data make analyses difficult, including (1) non-stationary within-day patterns; (2) substantial between-day heterogeneity, non-Gaussianity, and outliers; 3) large dimensionality due to the number of study participants, sleep periods, and time points. We evaluate and compare two methods: fast univariate inference (FUI) and functional additive mixed models (FAMM). We introduce a new approach for calculating p-values for testing a global null effect of covariates using FUI, and provide practical guidelines for speeding up FAMM computations, making it feasible for our data. While FUI and FAMM are philosophically different, they lead to similar point estimators in our study. In contrast to FAMM, FUI is fast, accounts for within-day correlations, and enables the construction of joint confidence intervals. Our analyses reveal that: (1) biguanide medication and sleep apnea severity significantly affect glucose trajectories during sleep, and (2) the estimated effects are time-invariant.
\end{abstract}


\section{Introduction}
Diabetes is a chronic disease characterized by elevated blood glucose levels and is recognized as a rising global epidemic of the 21st century. There are two main types of diabetes: type I results from the pancreas' inability to produce insulin; type II is characterized by insulin resistance, and insufficient amount of produced insulin. Diabetes is associated with considerable morbidity and mortality \citep{zimmet2001global}, and is linked to development of retinopathy \citep{sobrin2011candidate}, cardiovascular disease \citep{resnick2002diabetes}, lower extremity amputations \citep{moxey2011lower} and cognitive dysfunction \citep{kodl2008cognitive}. In 2019, approximately 463 million people worldwide had diabetes ($8.8\%$ of the adult population), with type II diabetes constituting about $90\%$ of the cases, and the rates are projected to rise \citep{saeedi2019global}. Therefore, it is important to understand the factors that affect the development of diabetes, and its progression with time.

Obstructive sleep apnea (OSA) is a sleep-related breathing disorder that is common among patients with type II diabetes, with an estimated $54-86\%$ prevalence range \citep{foster2009sleep, lam2010prevalence}. Multiple studies have reported associations between OSA, insulin resistance, and glucose intolerance \citep{Punjabi:2002, Lindberg:2012}. Despite this, the explicit effects of OSA severity on the glucose control of patients with type II diabetes are not well-understood.

Traditionally, glucose control is quantified by Hemoglobin A1c (HbA$_{1c}$), which is a measure of the long-term average glucose levels. However, the glucose profiles are highly non-linear and non-stationary, being sensitive to various environmental factors including quantity and type of meals, medications, physical activity, and stress levels. As HbA$_{1c}$ does not capture the within-night variability of glucose profiles, it cannot capture short term and dynamic associations between exposures, such as OSA severity, and blood glucose during sleep.

Continuous glucose monitors (CGMs) are small wearable medical devices that record the blood glucose levels at regular intervals throughout the day, with a typical interval being 5 minutes. Unlike HbA$_{1c}$, CGM provide detailed quantification of blood glucose levels during the entire 24-hour period, and thus play an increasing role in clinical practice and disease management \citep{rodbard2016continuous}.

\begin{figure}[!t]
    \centering
    \includegraphics[width=\textwidth]{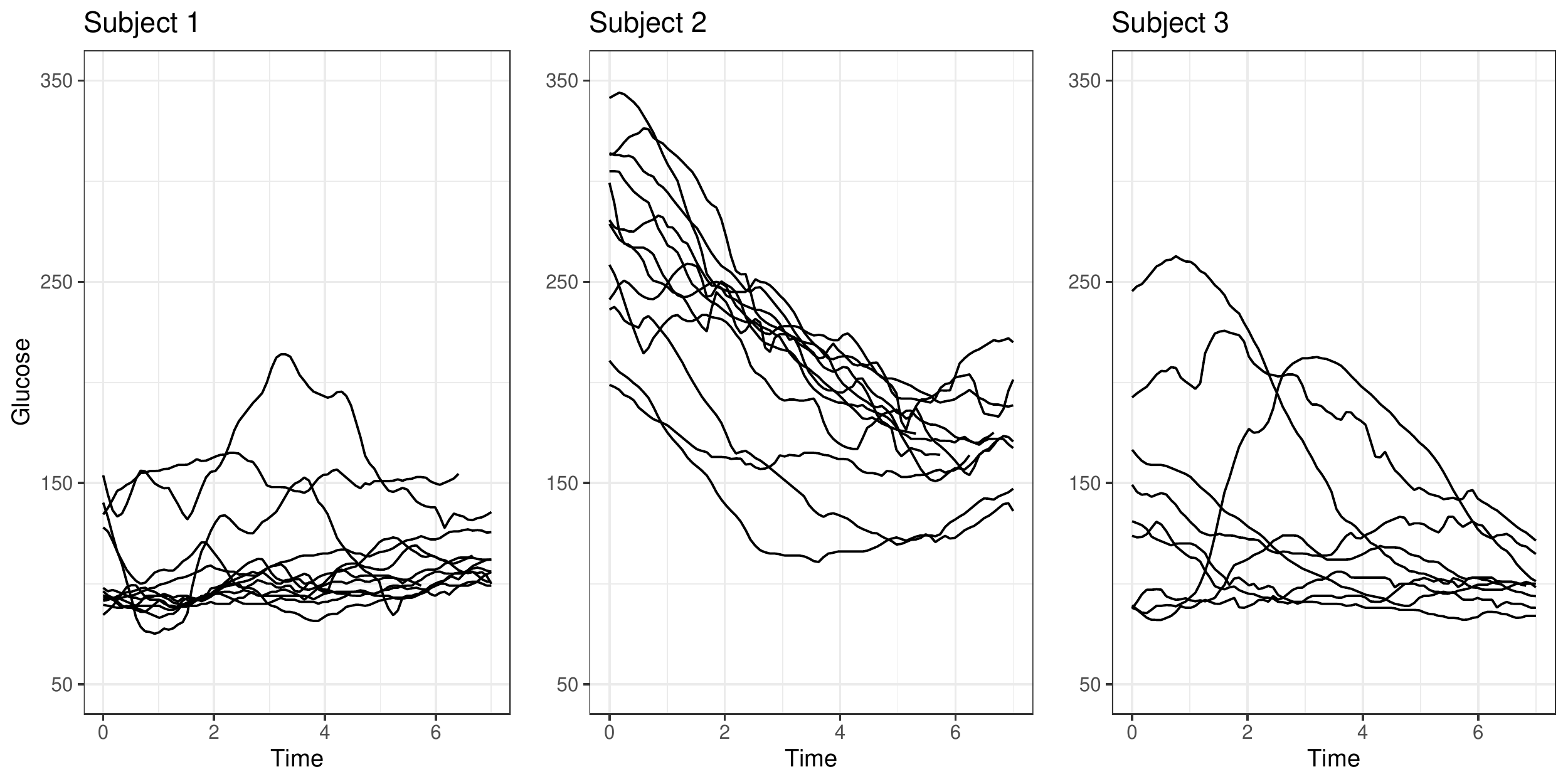}
    \caption{Glucose trajectories during sleep for three selected subjects. The x-axis is time from estimated sleep onset, one observation every five minutes. The glucose values are measured in mg/dL.}
    \label{fig:data}
\end{figure}

In this work, we aim to quantify the effects of OSA severity and medications on glucose control by analyzing CGM profiles of patients with type II diabetes from the HYPNOS study \citep{rooney2021rationale}. A unique characteristic of the study is the availability of concurrent data from CGM and wearable actigraphy devices. For the purpose of this study, actigraphy was used to estimate the sleep periods. Figure~\ref{fig:data} provides examples of profiles for three study participants during their estimated sleep periods. Time zero indicates estimated sleep onset time, and time on the x-axis is the time from the sleep onset. Focusing on sleep periods is important as: (1) OSA occurs during sleep; (2) sleep periods are less affected by confounding factors such as food intake or physical activity. However, when, what and how much the person ate before going to sleep is likely to affect both the blood glucose level at the beginning and its dynamics during the sleep period. A subset of these data was previously analyzed in \citet{Gaynanova:2022hy}. However this prior analysis focused on modeling variability using functional PCA  and did not specifically investigate the potentially time-varying fixed effects of covariates, such as OSA severity or diabetes medications, on glucose profiles.

It is common practice to extract summary measures (e.g., mean, standard deviation, coefficient of variation) from CGM data and use these summaries in subsequent analyses \citep{rodbard2016continuous,broll2021interpreting}. For example, prior study of the HYPNOS data \citep{aurora:2022} has used average glucose values over each sleep period and a linear mixed model to account for the multiple days of monitoring.  
Such approaches are easy to interpret as they use standard, well-known statistical models. The main disadvantage is that substantial information may be lost while compressing the CGM trajectory into a single number. Data compression may also make it impossible to test certain hypotheses of interest, e.g. one cannot ask or answer whether the effects of covariates are time-varying. 

In this paper we focus on modeling the complete measured glucose profile as a functional response. We are interested in how these profiles are associated with OSA severity and medication. The methodology used for analysis is in the general area of multilevel function-on-scalar regression \citep{bigelow2009bayesian,Di:2009dz,goldsmith2010, Greven:2010ik,morris2006wavelet}. While estimation in multilevel function-on-scalar regression has been well-studied, statistical inference remains an active area of research. In particular, the HYPNOS data have:  (1) large dimensionality with $174$ study participants, average of $10$ sleep periods per participant, and $84$ time points per period ($5$ minute intervals covering first seven hours from sleep onset); (2) highly non-stationary within-period patterns; (3) substantial between-period heterogeneity with non-Gaussianity and outliers. The main problems with existing methods for such data are: (1) computational feasibility and scalability; (2) availability of software; and (3) validity of resulting uncertainty estimates in the presence of within-curve functional correlations, outliers, and non-Gaussian errors. Our goal is to conduct inference on the HYPNOS data using computationally feasible methods that account for its multilevel functional structure and heterogeneity. We have identified only one method, the recently published Fast Univariate Inference (FUI) \citep{cui2022fast}, that can be adapted to achieve all of these goals. We contrast this approach with the traditional functional additive mixed models framework (FAMM) \citep{scheipl2015functional} as implemented in \textbf{refund} R package \citep{refundpackage}. 

Our major contribution is to compare FUI and FAMM in a realistic scenario with a new data type (CGM), providing illustration on the proper use of these methods and practical guidance on their implementation. We conclude that FUI is much faster than FAMM when accounting for the within-sleep period correlation structure of the data. Indeed, in our context fitting FAMM proved to be computationally prohibitive for the HYPNOS data. Moreover, even when FAMM does not account for within-sleep period correlations, the approach is very slow using default settings (over 12 hours on a standard laptop). We provide practical changes to the FAMM default settings that reduce computation time to only $7$ minutes. These changes may be known to FAMM experts, but most users would be unaware of where and how to adjust these settings. Our practical guidelines can substantially improve the user experience and, ultimately, the use of FAMM in practice. We also show that despite philosophical differences between the methods, FUI (accounting for within-period correlation) and FAMM (ignoring within-period correlation) lead to similar point estimators of fixed effects. The resulting FUI joint confidence bands are based on nonparametric case boostrap, and thus, unlike FAMM, take into account within-curve functional correlations, outliers and non-Gaussian distribution of the errors. While FAMM confidence bands are slightly narrower, they are pointwise rather than joint, and could not account for within-sleep period correlations. This comparison is new and was not investigated in the original FUI paper, which did not include simulations with correlated visit-specific data. 
Our second contribution is to extend FUI by quantifying the level of statistical significance of observed fixed effects (p-values for testing the null hypothesis of no effect) based on FUI joint confidence bands. To the best of our knowledge, this is the first time when joint confidence intervals are used to derive p-values for testing of functional effects. 

In summary, our case study illustrates how to practically address statistical challenges including: (1)~characterizing the association between covariates and glucose trajectories as a function of time from sleep onset; (2) constructing confidence intervals for these effects and providing measures of statistical significance of observed effects (p-values); (3) evaluating and comparing two published analytic methods (FUI AND FAMM); and (4) providing practical conclusions and advice for the use of these methods in future studies. Such case studies are crucial to the ultimate success of functional data analyses (FDA) approaches. Indeed, despite their success in the statistical literature, FDA is rarely used in practice. Successful case studies that go beyond typical didactic examples could help improve the acceptability of FDA methods in realistic scenarios.

The rest of the paper is organized as follows. Section~\ref{sec:data} describes data collection and processing and identifies specific analytic challenges. Section~\ref{sec::method} introduces the methodological approaches used and describes the new developments that complement the published results.  Section~\ref{sec:results} provides the results of the HYPNOS data set and explains their clinical implications. Section~\ref{sec:discuss} concludes with the discussion.

\section{Data Description}
\label{sec:data}
\subsection{Data collection}\label{subsec:collection}
The data analyzed in this paper was collected as part of the Hyperglycemic Profiles in Obstructive Sleep Apnea (HYPNOS) randomized clinical trial. The study population consisted of adults between  21 and 75 years old with type II diabetes and mild-to-severe OSA recruited from the community. The primary objective of the trial was to determine whether treatment with positive airway pressure (PAP) therapy is associated with improvements in glycemic measures. To investigate the effects of OSA severity and medication on the glucose control of patients with type II diabetes, we consider the data at the baseline visit (prior to randomization of study participants into control and PAP therapy groups).

The research protocol was approved by the Institutional Review Board on human research (Number: NA\_00093188). A detailed description of the trial protocol and implementation can be found in \citet{rooney2021rationale}. Below we provide a short summary.

Study participants were screened based on the point-of-care HbA$_{1c}$ measured with a DC Vantage Analyzer (Siemens Malvern PA), and a home sleep apnea test using Apnealink (Resmed San Diego, CA).  The oxygen desaturation index (ODI) was determined using the number of times the oxyhemoglobin decreased by 4\% per hour of sleep. Participants with HbA$_{1c}$ $\geq$ 6.5\% and ODI $\geq$ 5 events/hr were invited to enroll in the study. For each study participant, OSA severity was characterized as mild ($5\leq $ ODI $< 15$) or moderate-to-severe (ODI $\geq 15$). Exclusion criteria included pregnancy, any prior therapy for OSA, insulin use, change in glycemic medications in the previous six weeks, current oral steroid use, other sleep disorders, habitual sleep duration of $<$ 6 hours/night, any unstable medical condition. Study participants completed an actigraphy study using the Actiwatch (Philips Respironics, Murraysville, PA) and continuous glucose monitoring (CGM) using the Dexcom G4 Platinum sensor, which produces one measurement every 5 minutes. The actigraph was worn on the non-dominant wrist and the Dexcom sensor was placed $6$cm lateral to the umbilicus. Participants were instructed to wear both monitors for at least $7$ days and provide calibration glucose data for the Dexcom sensor twice a day according to the manufacturer's instructions.

\subsection{Extraction of glucose curves corresponding to sleep periods} \label{subsec:preprocess}
 Sleep periods were estimated using the actigraphy data and the proprietary algorithm of the Phillips Actiwatch wearable device. Actigraphy-estimated sleep periods that were shorter than 5 hours were excluded from the analysis. Only participants who had at least 5 sleep periods with concurrent CGM measurements were included in our analyses. This led to $1812$ actigraphy-estimated sleep periods for $174$ study participants ranging from $5$ to $20$ sleep periods per study participant, with a median of $11$ sleep periods. 
 We focus on the first 7 hours from the estimated sleep onset (time zero), and synchronize the measurement times across participants by linearly interpolating glucose trajectories at 5 min intervals from time zero. The interpolation interval matches the frequency of CGM device, and the interpolated trajectories are visually indistinguishable from the original ones.

Figure~\ref{fig:data} displays the glucose trajectories during the first seven hours of actigraphy-estimated
sleep for three study participants (selected to illustrate the variety of BG profiles observed in the
data). Each black solid line corresponds to one period of actigraphy-estimated sleep and time zero
corresponds to the actigraphy-estimated sleep onset.
 While typical glucose values range between 70 mg/dL and 120 mg/dL for people without diabetes, the spread of glucose values for patients with diabetes is much more variable, even during sleep. All three subjects in Figure~\ref{fig:data} exhibit high glucose values, with Subject 2 having the worst hyperglycemia with measurements in $[120, 350]$ mg/dL range. Furthermore, while the glucose values are expected to decrease during sleep due to the absence of food intake, this decreasing trend is not consistently observed across all subjects and sleep periods. Subject 1 has trajectories that tend to increase throughout the night, with several trajectories having peak in the middle of sleep. In contrast, all trajectories for Subject 2 are decreasing. The trajectories for Subject 3 are highly variable across nights. While most trajectories are decreasing, one trajectory starts at in-range values of 90 mg/dL, reaches hyperglycemic value of 210 mg/dL at 3 hours from sleep-onset, and goes back to in-range 100 mg/dL at 7 hours.

\subsection{Research questions and statistical challenges}\label{subsec:challenges}

In addition to the glucose trajectories, multiple covariates are available for each study participant: age, gender, BMI (coded as zero for for $\mbox{BMI}< 35$, and as one for $\mbox{BMI} > 35$), use of hypoglycemic medications (biguanides and sulfonylureas), point of care HbA$_{1c}$, and OSA severity status (mild OSA and moderate-to-severe OSA). 
Since point of care  HbA$_{1c}$ measurements were obtained prior to device placements, we use HbA$_{1c}$ as a baseline marker of diabetes severity. 

Our primary scientific goal is to investigate the effects of OSA severity and hypoglycemic medications on glucose trajectories during sleep after accounting for baseline HbA$_{1c}$, age, gender and BMI. Statistically, this corresponds to an inference question about the association between fixed scalar covariates (e.g., OSA severity, medication use) and functional responses (glucose trajectory during sleep). However, this is not a standard function-on-scalar regression (terminology introduced by \cite{reiss2010}) because the CGM exhibits: (1) highly non-stationary within-day patterns; (2) high between-day heterogeneity and  substantial departures from normality of the marginal distributions; (3) multi-level structure with large number of study participants ($174$) and time points ($84$). Our methodological goal is to conduct inference on these associations using computationally feasible methods that account for the known structure of the data.

\section{Methodology}
\label{sec::method}
We start by establishing the notation and model structure. Denote by $y_{ij}(t_k)$ the glucose measurement for subject $i=1,\ldots,I$, sleep period $j=1,\ldots,J_i$ (number of periods varies across study participants), at time $t_k$, $k=1,\ldots,K$, from sleep onset. In this application we use an equally spaced time index for the $7$ hour interval from sleep onset. The data structure is multilevel functional \citep{morris2006wavelet, bigelow2009bayesian, Di:2009dz, Greven:2010ik, SageBook, meyer2015bayesian} because multiple functions are observed for each study participant. Let $\mathbf{x}_i=(x_{i1},\ldots,x_{iR})^{\top}$ be the $R\times 1$ dimensional vector of covariates for participant $i$ and $X = (\mathbf{x}_1, \dots, \mathbf{x}_I)^{\top}$ be the $I\times R$ dimensional matrix, where each row contains the covariates for one study participant. We are interested in how these covariates affect the blood glucose trajectories from sleep onset.

To address this problem we consider the following multilevel function-on-scalar regression model
\begin{equation}
    y_{ij}(t) = \beta_0(t)  + \sum_{r=1}^R \beta_r(t) x_{ir}  + b_i(t) + \epsilon_{ij}(t)\;.
    \label{eq:model}
\end{equation}
Here $\beta_0(t)$ is the global intercept and could be interpreted as the average blood glucose level at time $t$ over study participants, $i$, and sleep periods, $j$, when covariates are equal to zero. The component $\beta_0(t)+\sum_{r=1}^R\beta_r(t) x_{ir}$ is the average fixed effect at time $t$ across visits and subjects with the covariates $\mathbf{x}_i$. The functions, $\beta_r(\cdot)$ for $r=0,\ldots,R$, are assumed to be continuous and smooth. The random intercept $b_i(t)$ is the study participant specific deviation from the population mean at time $t$. It plays a similar role to that of random intercepts in standard linear mixed effects models. The major difference in our context is that $b_i(\cdot)$ is multivariate and smooth. We assume that $b_i(\cdot)\sim N(\mathbf{0}_K,\mathbf{\Sigma}_{b,K})$, mutually independent, where $\mathbf{0}_K$ is the $K\times 1$ dimensional vector of zeros, and $\mathbf{\Sigma}_{b,K}$ is the $K\times K$-dimensional between-study-participant covariance matrix after accounting for fixed effects. We also assume that $\epsilon_{ij}(\cdot)\sim N(\mathbf{0}_K,\mathbf{\Sigma}_{\epsilon,K})$, where $\mathbf{\Sigma}_{\epsilon,K}$ is the $K\times K$ dimensional within-study-participant covariance matrix. Note that it would be too restrictive to assume that $\epsilon_{ij}(t)$ are uncorrelated across $t$ because model~(\ref{eq:model}) would not be generative for the type of data observed in the HYPNOS study. Indeed, such a model would just produce white noise around the study participant specific mean. In the actual data the period-specific deviations from the study-participant-specific means are highly structured and smooth. 

We consider two competing methods for estimation and inference of model~(\ref{eq:model}): fast univariate inference (FUI) \citep{cui2022fast} and  functional additive mixed model (FAMM) \citep{scheipl2015functional}. The two approaches are philosophically different. FUI fits univariate mixed models at each time point and then uses a nonparametric case bootstrap of study participants to conduct inference. FAMM builds a joint model for the fixed and random functional effects and uses mixed effects modeling for inference. Doing this raises substantial computational problems, especially when one needs to model both the study participant-specific and period-specific effects. In fact, we could not fit FAMM on our data for a model that includes the structured period-specific effects. Consequently, we fit FAMM using a simplified model, where we assume that the residuals $\epsilon_{ij}(t)$ are independent across time. Both methods provide similar point estimators, though FUI provides slightly wider joint confidence intervals compared to pointwise intervals from FAMM. As data sets get larger, the computational advantage of FUI is likely to increase, at least given the current implementation of FAMM.  While both methods are described in detail in the respective papers, for presentation completeness we provide a summary of FUI in Section~\ref{subsec:fui} and of FAMM in Section~\ref{subsec:famm}, with additional practical guidelines.

\subsection{Estimation and inference with FUI}
\label{subsec:fui}
FUI fits separate linear mixed models at each time point. Therefore, it does not require the independence of residuals across the time domain. To be specific, for each $t$ from the discrete set $\{t_1, \dots, t_K \}$, FUI fits a univariate mixed effects model using, for example, \textbf{lme4} in the \textbf{R}  \citep{lme4package}. From the pointwise univariate models, we obtain estimators of the coefficient vectors $\bm{\hat \beta}_r = (\hat \beta_r(t_1), \dots, \hat \beta_r(t_K))$. These coefficients can then be smoothed; see, for example, \citet{cui2022fast}. 

To conduct fixed-effects inference,  a bootstrap of study-participants is used. That is, for a bootstrap $b=1,\ldots,B$, study participants $\{1, \dots, I\}$ are sampled with replacement and estimators of the effects of interest $\{\bm{\hat \beta}^{(b)}_r\}$ are obtained. For each effect of interest $r=1,\ldots,R$, we obtain the mean,  $\bm{\hat \beta}_r$, and covariance, $\text{var}(\bm{\hat \beta_r})$, from the $B$ bootstrap estimators.  The approach proceeds by sampling $\bm{\hat \beta}_{r}^{(n)} \sim \mathcal{N}\{\bm{\hat\beta}_r, \text{var}(\bm{\hat \beta}_r)\}$ $N$ times and calculating the distribution of the standardized maximum deviations $u_n = \max |\bm{\hat\beta}_{r}^{(n)} - \bm{\hat\beta_r}| / \sqrt{\text{diag}(\text{var}\{\bm{\hat \beta}_r)\}}$. The empirical $1-\alpha$ quantile $q_{1-\alpha}$ of the distribution $\{u_1, \dots, u_N \}$ is used to compute the joint confidence bands for the effects as $\bm{\hat\beta}_r \pm q_{1-\alpha}\sqrt{ \text{diag}(\text{var}(\bm{\hat\beta}_r))}$. Simulating from the multivariate normal distribution may be difficult when the dimension of the functional domain is very large. In this situation, \citet{cui2022fast} propose to use principal component decomposition of the $B\times K$-~dimensional matrix obtained by row binding the estimators $\{\bm{\hat \beta}^{(b)}_r\}$. This is not necessary in our application as the dimension of the covariance operator is $84\times 84$.

FUI implicitly allows for time-dependent error curves, $\epsilon_{ij}(\cdot)$, because fitting is done separately at every time point. Correlation of the errors is accounted for when calculating the confidence intervals based on the bootstrap of study participants. This is a simple but highly effective work-around for direct modeling of highly complex covariance structures. It does not impact the pointwise confidence intervals but provides a way of calculating pointwise and joint confidence intervals. In turn, joint confidence intervals can be used for global testing of parameter coefficients that account for complex dependence structures.  Moreover, this approach allows for more flexible modeling of the covariance including latent subgroups (e.g., when certain subgroups have a different dependence structure). While FUI is a very powerful approach to inference, the original paper \citep{cui2022fast} does not provide (1) an assessment of the method in the presence of within-period correlations (which is the case for our HYPNOS data); (2) a one-number summary (e.g., p-value) for testing the null hypotheses about the functional effects, $\beta_r(\cdot)$.

\subsubsection{Obtaining the p-value for testing the fixed effects}\label{subsubsec:testing}
FUI provides a way of constructing joint confidence intervals for $\beta_r(\cdot)$ at any level $\alpha$. However, it does not provide a one-number summary (e.g., p-value) for testing the null hypothesis about the functional effects. Specifically, we are interested in testing:
\begin{align*}
    H_0:~&\beta_r(t)=0\;\;\textrm{for\;every} \;t\in S\;, \\
    H_A:~&\beta_r(t)\neq 0\;\;\textrm{for\;at\;least\;one} \;t\in S\;.
\end{align*}
Here $S$ is any subset of the domain of the function $\beta_r(\cdot)$. In our example $S\subset [0,7]$, because we consider the first $7$ hours from sleep onset. It is typical to use the whole interval, i.e. $S= [0,7]$, but any other subset of $[0, 7]$ could be considered. A level $\alpha$ test rejects the null hypothesis if there exists at least one $t\in S$ for which the $1-\alpha$ joint confidence interval around $\beta_r(t)$ does not contain zero.

We propose to calculate the p-value for this test as the smallest level $\alpha$ for which at least one of the confidence intervals around $\beta_r(t)$ for $t\in S$ does not contain zero. This approach is provided for testing equality with zero. However, the approach can easily be extended to testing any null hypothesis of the type 
\[H_0:\beta_r(t)=\beta_{0r}(t)\;\;\textrm{for\;every} \;t\in S\;\]
using exactly the same computation, where the function $\beta_{0r}(\cdot) \equiv 0$ is replaced by another function $\beta_{0r}(\cdot)$. An advantage of having the p-value is that it provides an interpretable, quick summary of the evidence against the null hypothesis which complements the visual inspections of joint confidence intervals. Throughout this paper we are using these newly proposed p-values. To the best of our knowledge, this is the first time a p-value based on joint confidence intervals for a functional coefficient is proposed.

\subsection{Estimation and inference with FAMM}
\label{subsec:famm}
The functional additive mixed model (FAMM) expands each of the $r = 0, \dots, R$ fixed effect coefficient functions, $\beta_r(\cdot)$, in model~\eqref{eq:model} using a spline basis  $\beta_r(t) = \sum_{j=1}^S c_{rj}f_{j}(t)$. The type and number of basis functions is kept the same for all fixed effects, except for the intercept term, which uses a larger number of basis functions. This is the strategy advocated by \citet{ruppert2003semiparametric}. A quadratic penalty is imposed to smooth the spline coefficients, and the approach allows the use of different penalties for each fixed effect coefficient. 

For the subject-specific random intercept $b_i(\cdot)$ and period-specific time-dependent $\epsilon_{ij}(\cdot)$, FAMM uses the tensor-product expansion, more details for which are provided in \citet{scheipl2015functional}  and \citet{wood2017generalizedbook}. The tensor-product uses a Kronecker product to cross the bases for the random and time-dependent components. This tensor-product approach, however, creates a significant computational bottleneck as the total number of basis functions is a product between the number of participants, the number of visits, and the number of the basis functions for the time component. Indeed, with $10$ basis functions for the time component and $174$ participants, the basis for the random effects consists of $1,740$ functions. If we include the period-specific time-dependent noise, with $1,812$ visits, we end up with a basis of $18,120$ functions for the noise term. 

For these reasons,  it is impossible to fit the full model~(\ref{eq:model}) using FAMM on our data set. Instead, we will fit FAMM on a model that assumes that the residuals, $\epsilon_{ij}(\cdot)$, are independent across time. This model ignores the strong  within-sleep period structure of the data (see Figure~\ref{fig:data}), but it is the only FAMM model we could fit on our data.

The confidence intervals produced by FAMM are pointwise based on Bayesian credible intervals \citep{scheipl2015functional, wood2017generalizedbook}. To produce joint confidence intervals that account for within-sleep period structure, one could theoretically run a case bootstrap similar to the one used for FUI. However, FAMM is already quite slow and re-fitting it multiple times would be computationally prohibitive.

\subsubsection{Fitting FAMM in R}
\label{subsec:famm_implementation}

To fit FAMM, we use the \textbf{pffr} function in the R  package \textbf{refund} \citep{refundpackage}. The underlying model uses the generalized additive model (GAM) implemented in the package \textbf{mgcv} \citep{mgcvpackage}. The FAMM model~\eqref{eq:model} that accounts for within-sleep period correlations has the following syntax:

\begin{verbatim}
    fit.slow = pffr(y ~ age + sex + biguanide + sulfonylurea + 
                        hba1c + bmi + osa + s(id, bs="re") 
                        + s(idperiod, bs="re"), 
                        data = CGMdata, algorithm = "gam", method = "REML", 
                        bs.yindex = list(bs = "ps", k = 10, m = c(2, 1)),
                        bs.int = list(bs = "ps", k = 30, m = c(2, 1)))
\end{verbatim}
The \verb!CGMdata! is a data frame with $1,812$ rows corresponding to individual sleep periods. The columns contain values of $7$ fixed covariates (\verb!age!, \verb!sex!, \verb!biguanide!, \verb!sulfonylurea!,  \verb!hba1c!, \verb!bmi!, \verb!osa!), indicator of study participant (\verb"id" with $174$ levels), indicator of sleep period (\verb!idperiod! with $1,812$ levels), and a vector of glucose values (\verb!y! of length $84$). Covariates enter the model using standard \verb"pffr" syntax, with  \verb!~x_r! specifying a linear function effect of covariate \verb!x_r! that varies smoothly over time (term $\beta_r(t)x_{ir}$ in \eqref{eq:model}).  The term \verb!s(id, bs="re")! incorporates the study participant-specific random intercept, $b_i(\cdot)$. The term \verb!s(idperiod, bs="re")! incorporates the period-specific random effect. 
The parameters for the type and number of basis functions are specified in the \verb!bs.yindex! for the covariates and \verb!bs.int! for the global intercept term. For all covariates we used P-splines (\verb!bs="ps"!) with $10$ knots (\verb!k=10!) and second order basis function with first-order difference penalty (\verb!m=c(2,1)!). For the intercept we used P-splines (\verb!bs="ps"!) with $30$ knots (\verb!k=30!) and second order basis function with first-order difference penalty (\verb!m=c(2,1)!). A summary of the resulting fit can be obtained by running \verb!summary(fit.pffr.slow)!. For other basis options, see \citet{wood2017generalizedbook}.

Unfortunately, fitting this model proved to be computationally infeasible on the HYPNOS data. The high computational cost can be attributed to three factors: (i) modeling of sleep period-specific effects via \verb!s(idperiod,bs="re")!; (ii) use of default \verb!algorithm = "gam"! with $174$ levels in \verb!s(id,bs="re")!; (iii) presence of continuous covariates (age, HbA$_{1c}$). To circumvent these difficulties, we: (i) fit a model without the sleep period-specific effects, that is delete the term \verb!s(idperiod,bs="re")!; (ii) change \verb!algorithm = "gam"! to \verb!algorithm = "bam"! (which is recommended for larger data sets) and (iii) activate the option \verb"discrete=TRUE" in \verb"bam" (default is FALSE). All these changes proved essential to run the model, even after ignoring sleep period-specific correlations. The updated R syntax is

\begin{verbatim}
    fit.fast = pffr(y ~ age + sex + biguanide + sulfonylurea + 
                        hba1c + bmi + osa + s(id, bs="re"), 
                        data = CGMdata, algorithm = "bam",
                        method = "fREML", discrete = TRUE,
                        bs.yindex = list(bs = "ps", k = 10, m = c(2, 1)),
                        bs.int = list(bs = "ps", k = 30, m = c(2, 1)))
\end{verbatim}

Dropping the \verb!s(idperiod, bs="re")! term leads to model miss-specification. More precisely, the fit is obtained under the assumption of independence of within-sleep period deviations from the study participant-specific mean. As we will show, this does not seem to impact the point estimators, though it has an effect on the length of confidence intervals and their potentival validity in terms of coverage. Using \verb!bam! instead of \verb"gam" further reduces computational complexity by evaluating the basis functions on a random subset of the data and iteratively maximizing the likelihood \citep{wood2015generalized}. Setting \verb!discrete = TRUE! in  \verb!bam! allows to discretize continuous covariates and reduce the computational complexity by reducing the memory load \citep{wood2017generalized}. Neither the original FAMM model (that incorporated sleep period-specific effects), nor the adjusted FAMM model (that ignored sleep period-specific effects) could be fit on our data with default algorithm settings (for both models, the computations were interrupted after $12$ hours). After using the \verb!bam! and \verb"discrete=TRUE" options, the misspecified model that ignored sleep period-specific effects ran in $7$ minutes. One could think about these details as ``computational tricks". However, we posit that they are important components that are highlighted by complex applications, such as ours. This is the reason why testing methods and software should be done by multiple research groups and under realistic scenarios. We posit that one of the most important contribution of our paper is to test if and how state-of-the-art methods perform in realistic scenarios.

\section{Results}\label{sec:results}


Our first question of interest is investigating the effects of OSA severity and hypoglycemic medications on glucose trajectories during sleep after accounting for baseline HbA$_{1c}$, age, gender and BMI. BMI is coded as zero for for $\mbox{BMI}< 35$, and as one for $\mbox{BMI} > 35$. 
OSA severity is coded as zero for mild OSA, and as one for moderate-to-severe OSA. For hypoglycemic medications, we consider the biguanide family (coded as $0$ for no use, and $1$ for active use), and sulfonylurea family (coded as $0$ for no use, and $1$ for active use).  Both biguanides and sulfonylureas are families of oral drugs commonly prescribed for type II Diabetes to reduce glucose levels. Metformin is the most recognised drug in the biguanides group and is commonly used as a first-line treatment for type II diabetes. While biguanides work by suppressing the production of glucose in the liver, sulfonylureas promote the body's production of insulin. The two families of medications are not mutually exclusive. Out of 174 patients, 18 take neither type of the drug, 90 take biguanide only, 11 take sulfonylurea only, and 55 take both drugs. Sex is modeled as a binary variable with $1$ corresponding to males, and $0$ to females. Age and HbA$_{1c}$ are modeled as continuous variables.


\begin{figure}[!t]
    \centering
    \includegraphics[width = \textwidth]{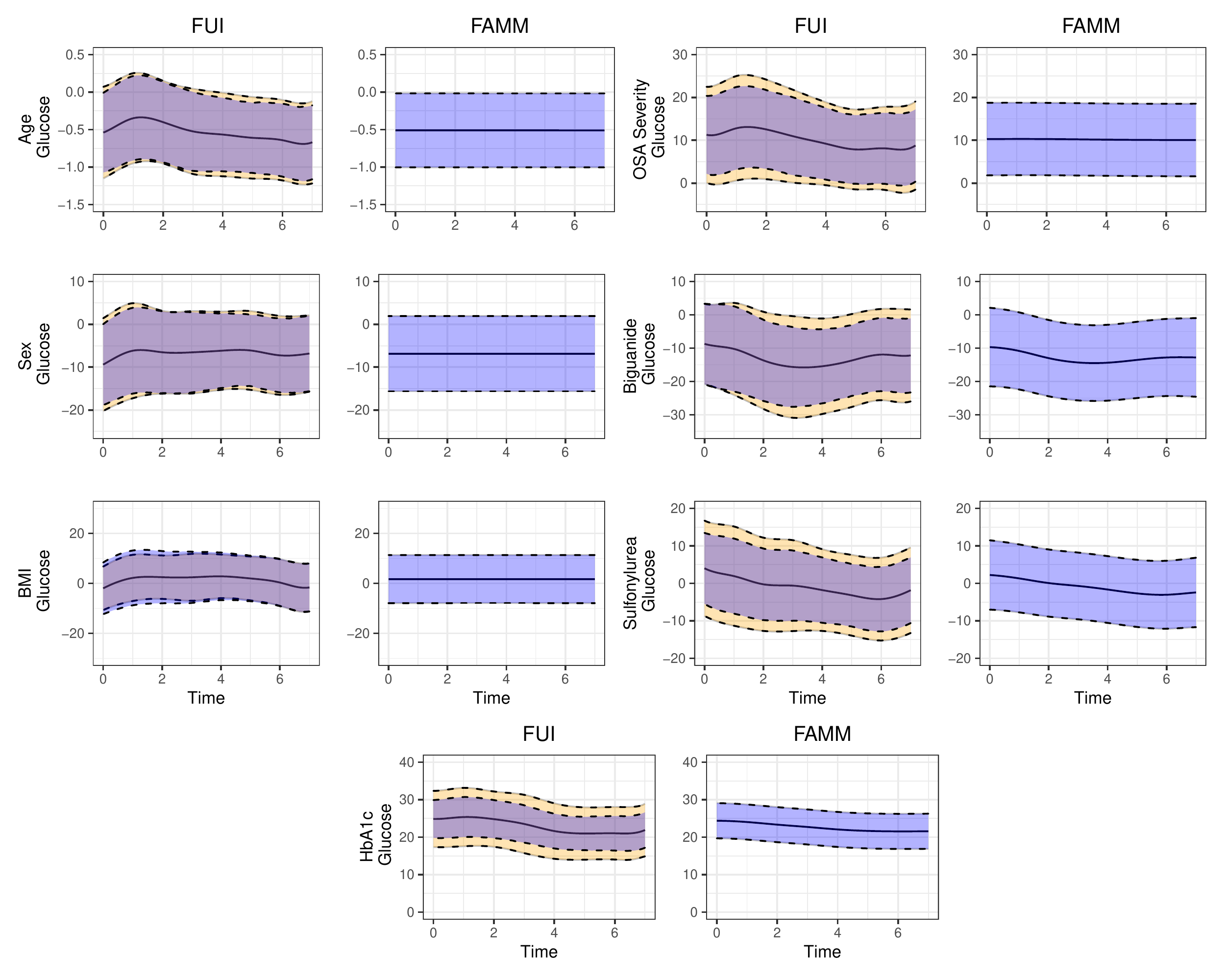}
    \caption{Estimated coefficient functions together with the 95\%
    confidence intervals from fast univariate inference method (FUI) and functional additive mixed model (FAMM) described in Sections~\ref{subsec:fui} and \ref{subsec:famm}. For FAMM, the confidence intervals are pointwise. For FUI, both pointwise and joint confidence intervals are displayed, 
    with joint intervals obtained as in Section~\ref{subsec:fui}.}
    \label{fig:covs}
\end{figure}

We apply both FAMM and FUI methods to conduct inference on fixed effect coefficients $\beta_r(t)$ in model~\eqref{eq:model}. Figure~\ref{fig:covs} displays the estimated coefficient curves, $\hat \beta_r(t)$, for each covariate, together with the corresponding $95\%$ confidence bands. It is reassuring that the point estimators for FUI and FAMM are relatively close, though some differences can be observed. In particular, FUI estimates are, in general, more variable across time. For FAMM, we only display the pointwise confidence bands obtained under the misspecified model that does not account for the sleep period-specific correlation. For FUI, we display both the pointwise and joint confidence bands (obtained as described in Section~\ref{subsec:fui}). FUI pointwise bands are on average 3\% wider than FAMM across all covariates with the minimum of 1\% (for biguanide and sulfonylurea), and the maximum of 4\% (for all remaining covariates). FUI joint bands are on average 22\% wider than FAMM, with the minimum of 11\% (for Sex) and a maximum of 57\% (for HbA$_{1c}$). Somewhat surprisingly, the FUI joint band for BMI is shorter than both FUI pointwise band and FAMM joint band. We suspect that this discrepancy is due to differences in calculation of standard errors. FUI pointwise bands use standard errors from pointwise mixed models (corresponding to unsmoothed coefficients), whereas the joint bands use standard errors from bootstrap, where at each replication the coefficients are smoothed. In case the true underlying coefficient is very smooth, this can lead to smaller standard errors from bootstrap.

 The p-values based on FUI joint confidence bands are provided in Table~\ref{table:testing}. To the best of our knowledge, this is the first time when such p-values are shown. We expect that they will become popular because they are intuitive. Both BMI and Sex were not significant at the level $\alpha=0.05$. As expected, HbA$_{1c}$ is very strongly associated with the CGM curves (p-value$<$0.001). This makes sense, as HbA$_{1c}$ is thought to measure an average of glucose values over the past 3 months \citep{nathan2007}.
 The FUI analysis seems to indicate that this association is stronger in the first part of the sleep period, though the association is statistically significant across the entire sleep period.  Results indicate that patients with higher baseline HbA$_{1c}$ have overall higher glucose levels. While the effect of sulfonylurea medication on glucose profiles during sleep is not significant (p-value$=$0.673), the effect of biguanide is (p-value$=$0.033). Specifically, biguanide is associated with lower glucose values during sleep after accounting for HbA$_{1c}$. Panels labeled biguanide in Figure~\ref{fig:covs} indicate that the effect is the strongest approximately $3$ hours after the sleep onset. A possible explanation is that biguanide is typically taken with dinner in the evening and the effect of the drug is delayed. Age is also significant (p-value=$0.008$), with higher age being associated with lower glucose values. One possible explanation is that older patients may have more experience in managing their diabetes. OSA severity has a significant negative effect on blood glucose even after accounting for HbA$_{1c}$, with larger glucose values during sleep for patients with moderate-to-severe OSA compared to patients with mild OSA.

\begin{table}[!t]
\centering
\caption{P-values for each fixed effect computed using the method in Section~\ref{subsubsec:testing}}
\label{table:testing}
\begin{tabular}{lc}\\\toprule  
Covariates & P-values \\\midrule
Age & $0.008$\\  \midrule
Sex & $0.099$\\  \midrule
BMI & $0.832$\\  \midrule  
OSA Severity & $0.026$\\  \midrule
Biguanide & $0.033$\\  \midrule
Sulfonylurea & $0.673$\\ \midrule
HbA$_{1c}$ & $<0.001$\\
\bottomrule
\end{tabular}
\end{table}

Overall, our empirical results highlight that FAMM and FUI methods result in similar width of the confidence bands and similar estimates of the coefficients functions with FUI providing only slightly wider joint confidence intervals. Further, we note that FUI estimates are, in general, more variable across time. Both FUI and FAMM are powerful estimation approaches that are designed to work with complex dependencies in the functonal data. In practice, however, fitting FAMM may be computationally challenging and may need case-by-case adjustment of the model specifications. Based on 10 runs, we report the average fitting time for both methods on Intel Core i7. FUI takes \textit{less than two seconds} to fit pointwise linear mixed models on the whole data, whereas FAMM with the options selected in Section~\ref{subsec:famm_implementation} takes \textit{seven minutes}. At the same time, without computation speed-up, FAMM takes more than 12 hours (the computations were interrupted at 12 hours). Combining FUI with case bootstrap to perform inference leads to overall computation time of two minutes with 100 bootstrap replications performed sequentially, which is still at least 3-fold less than the one time fit of FAMM.

\section{Discussion}\label{sec:discuss}

In this paper, we consider the problem of inference in multilevel function-on-scalar regression framework using two distinct methodologies, FAMM and FUI, in the context of glucose trajectories measured by CGM during sleep. Our results indicate that the glucose levels during sleep are significantly higher in patients with moderate-to-severe OSA compared to patients with mild OSA. Since heightened glucose levels in type II diabetes are associated with major adverse health effects, our findings suggest that OSA treatment options may need to be considered as part of overall diabetes management plan in addition to traditional diet and exercise interventions. We have also found significant effects of participants' age, baseline diabetes severity as measured by HbA$_{1c}$, and biguanide medication. 

In terms of methods comparison, we found that FUI outperforms FAMM in terms of the fitting time and allows changes in model complexity without large losses in computational time. Indeed, fitting the full model~(\ref{eq:model}) with FAMM is computationally prohibitive on our data without further methodological developments. When fitting the simplified FAMM model, FUI and FAMM provide similar estimates of the fixed effects. However, the FUI confidence bands are slightly wider as they are joint and account for within-period correlations, whereas the FAMM confidence bands are pointwise and are based on a misspecified model. We also provide an important improvement for FUI: computation of p-values for testing the joint fixed effects using joint confidence intervals at different $\alpha$ levels. As the proposed p-values are intuitive we expect that they will become popular in quantifying the significance of functional fixed effects in other applications.

There are multiple opportunities for further methodological and scientiﬁc research. First, the joint inference across the time domain is a unique feature of FUI, which FAMM currently does not support. To perform joint inference, FUI relies on the bootstrap of the study participants, which is computationally efficient given its low model fitting cost. In contrast, fitting FAMM, even for a misspecified model, requires considerable computational resources, which reduces the appeal of the bootstrap. 
Second,  while we focused on inference for fixed effects, subject-specific inference could be of substantial interest in the context of personalized assessment of glucose control during sleep. The main difficulties for subject-specific inference are: (1) substantial between-subject variability; and (2) small number of sleep periods per subject. 
Third, it may be important to monitor dynamically individual glucose trajectories and identify early unusual patterns that could be predictive of adverse health outcomes (e.g., hyper- or hypo-glycemia).

\bibliographystyle{biom}
\bibliography{references}

\begin{thebibliography}{}

\bibitem[\protect\citeauthoryear{Aurora, Gaynanova, Patel, and Punjabi}{Aurora
  et~al.}{2022}]{aurora:2022}
Aurora, R.~N., Gaynanova, I., Patel, P., and Punjabi, N.~M. (2022).
\newblock Glucose profiles in obstructive sleep apnea and type 2 diabetes
  mellitus.
\newblock {\em Sleep Medicine} page in press.

\bibitem[\protect\citeauthoryear{Bates, M{\"a}chler, Bolker, and Walker}{Bates
  et~al.}{2015}]{lme4package}
Bates, D., M{\"a}chler, M., Bolker, B., and Walker, S. (2015).
\newblock Fitting linear mixed-effects models using {lme4}.
\newblock {\em Journal of Statistical Software} {\bf 67,} 1--48.

\bibitem[\protect\citeauthoryear{Bigelow and Dunson}{Bigelow and
  Dunson}{2009}]{bigelow2009bayesian}
Bigelow, J.~L. and Dunson, D.~B. (2009).
\newblock Bayesian semiparametric joint models for functional predictors.
\newblock {\em Journal of the American Statistical Association} {\bf 104,}
  26--36.

\bibitem[\protect\citeauthoryear{Broll, Urbanek, Buchanan, Chun, Muschelli,
  Punjabi, and Gaynanova}{Broll et~al.}{2021}]{broll2021interpreting}
Broll, S., Urbanek, J., Buchanan, D., Chun, E., Muschelli, J., Punjabi, N.~M.,
  and Gaynanova, I. (2021).
\newblock Interpreting blood glucose data with r package iglu.
\newblock {\em PloS one} {\bf 16,} e0248560.

\bibitem[\protect\citeauthoryear{Crainiceanu, Caffo, and Morris}{Crainiceanu
  et~al.}{2013}]{SageBook}
Crainiceanu, C., Caffo, B., and Morris, J. (2013).
\newblock Multilevel functional data analysis.
\newblock In {\em The SAGE handbook of multilevel modeling}, pages 223--248.
  SAGE Publications Ltd, London.

\bibitem[\protect\citeauthoryear{Cui, Leroux, Smirnova, and Crainiceanu}{Cui
  et~al.}{2022}]{cui2022fast}
Cui, E., Leroux, A., Smirnova, E., and Crainiceanu, C.~M. (2022).
\newblock Fast univariate inference for longitudinal functional models.
\newblock {\em Journal of Computational and Graphical Statistics} {\bf 31,}
  219--230.

\bibitem[\protect\citeauthoryear{Di, Crainiceanu, Caffo, and Punjabi}{Di
  et~al.}{2009}]{Di:2009dz}
Di, C.-Z., Crainiceanu, C., Caffo, B., and Punjabi, N. (2009).
\newblock {Multilevel functional principal component analysis}.
\newblock {\em Annals of Applied Statistics} {\bf 3,} 458 -- 488.

\bibitem[\protect\citeauthoryear{Foster, Sanders, Millman, Zammit, Borradaile,
  Newman, Wadden, Kelley, Wing, Pi~Sunyer, Darcey, Kuna, and for~the Sleep
  AHEAD Research~Group}{Foster et~al.}{2009}]{foster2009sleep}
Foster, G.~D., Sanders, M.~H., Millman, R., Zammit, G., Borradaile, K.~E.,
  Newman, A.~B., Wadden, T.~A., Kelley, D., Wing, R.~R., Pi~Sunyer, F.~X.,
  Darcey, V., Kuna, S.~T., and for~the Sleep AHEAD Research~Group (2009).
\newblock Obstructive sleep apnea among obese patients with type 2 diabetes.
\newblock {\em Diabetes Care} {\bf 32,} 1017--1019.

\bibitem[\protect\citeauthoryear{Gaynanova, Punjabi, and Crainiceanu}{Gaynanova
  et~al.}{2022}]{Gaynanova:2022hy}
Gaynanova, I., Punjabi, N., and Crainiceanu, C. (2022).
\newblock {Modeling continuous glucose monitoring (CGM) data during sleep}.
\newblock {\em Biostatistics} {\bf 23,} 223--239.

\bibitem[\protect\citeauthoryear{Goldsmith, Zipunnikov, and Schrack}{Goldsmith
  et~al.}{2015}]{goldsmith2010}
Goldsmith, A., Zipunnikov, V., and Schrack, J. (2015).
\newblock Generalized multilevel function-on-scalar regression and principal
  component analysis.
\newblock {\em Biometrics} {\bf 71,} 344--353.

\bibitem[\protect\citeauthoryear{Goldsmith, Scheipl, Huang, Wrobel, Di, Gellar,
  Harezlak, McLean, Swihart, Xiao, Crainiceanu, and Reiss}{Goldsmith
  et~al.}{2020}]{refundpackage}
Goldsmith, J., Scheipl, F., Huang, L., Wrobel, J., Di, C., Gellar, J.,
  Harezlak, J., McLean, M.~W., Swihart, B., Xiao, L., Crainiceanu, C., and
  Reiss, P.~T. (2020).
\newblock {\em refund: Regression with Functional Data}.
\newblock R package version 0.1-23.

\bibitem[\protect\citeauthoryear{Greven, Crainiceanu, Caffo, and Reich}{Greven
  et~al.}{2010}]{Greven:2010ik}
Greven, S., Crainiceanu, C.~M., Caffo, B., and Reich, D. (2010).
\newblock {Longitudinal functional principal component analysis}.
\newblock {\em Electronic Journal of Statistics} {\bf 4,} 1022 -- 1054.

\bibitem[\protect\citeauthoryear{Kodl and Seaquist}{Kodl and
  Seaquist}{2008}]{kodl2008cognitive}
Kodl, C.~T. and Seaquist, E.~R. (2008).
\newblock Cognitive dysfunction and diabetes mellitus.
\newblock {\em Endocrine reviews} {\bf 29,} 494--511.

\bibitem[\protect\citeauthoryear{Lam, Lui, Lam, Ong, Lam, and Ip}{Lam
  et~al.}{2010}]{lam2010prevalence}
Lam, D.~C., Lui, M.~M., Lam, J.~C., Ong, L.~H., Lam, K.~S., and Ip, M.~S.
  (2010).
\newblock Prevalence and recognition of obstructive sleep apnea in chinese
  patients with type 2 diabetes mellitus.
\newblock {\em Chest} {\bf 138,} 1101--1107.

\bibitem[\protect\citeauthoryear{Lindberg, Theorell-Haglöw, Svensson,
  Gislason, Berne, and Janson}{Lindberg et~al.}{2012}]{Lindberg:2012}
Lindberg, E., Theorell-Haglöw, J., Svensson, M., Gislason, T., Berne, C., and
  Janson, C. (2012).
\newblock {Sleep Apnea and Glucose Metabolism A Long-term Follow-up in a
  Community-Based Sample}.
\newblock {\em Chest} {\bf 142,} 935--942.

\bibitem[\protect\citeauthoryear{Meyer, Coull, Versace, Cinciripini, and
  Morris}{Meyer et~al.}{2015}]{meyer2015bayesian}
Meyer, M.~J., Coull, B.~A., Versace, F., Cinciripini, P., and Morris, J.~S.
  (2015).
\newblock Bayesian function-on-function regression for multilevel functional
  data.
\newblock {\em Biometrics} {\bf 71,} 563--574.

\bibitem[\protect\citeauthoryear{Morris and Carroll}{Morris and
  Carroll}{2006}]{morris2006wavelet}
Morris, J.~S. and Carroll, R.~J. (2006).
\newblock Wavelet-based functional mixed models.
\newblock {\em Journal of the Royal Statistical Society: Series B (Statistical
  Methodology)} {\bf 68,} 179--199.

\bibitem[\protect\citeauthoryear{Moxey, Gogalniceanu, Hinchliffe, Loftus,
  Jones, Thompson, and Holt}{Moxey et~al.}{2011}]{moxey2011lower}
Moxey, P., Gogalniceanu, P., Hinchliffe, R., Loftus, I., Jones, K., Thompson,
  M., and Holt, P. (2011).
\newblock Lower extremity amputations—a review of global variability in
  incidence.
\newblock {\em Diabetic Medicine} {\bf 28,} 1144--1153.

\bibitem[\protect\citeauthoryear{Nathan, Turgeon, and Regan}{Nathan
  et~al.}{2007}]{nathan2007}
Nathan, D., Turgeon, H., and Regan, S. (2007).
\newblock Relationship between glycated haemoglobin levels and mean glucose
  levels over time.
\newblock {\em Diabetologia} {\bf 50,} 2239--2244.

\bibitem[\protect\citeauthoryear{Punjabi, Sorkin, Katzel, Goldberg, Schwartz,
  and Smith}{Punjabi et~al.}{2002}]{Punjabi:2002}
Punjabi, N.~M., Sorkin, J.~D., Katzel, L.~I., Goldberg, A.~P., Schwartz, A.~R.,
  and Smith, P.~L. (2002).
\newblock {Sleep-disordered Breathing and Insulin Resistance in Middle-aged and
  Overweight Men}.
\newblock {\em American Journal of Respiratory and Critical Care Medicine} {\bf
  165,} 677--682.

\bibitem[\protect\citeauthoryear{Reiss, Huang, and Mennes}{Reiss
  et~al.}{2010}]{reiss2010}
Reiss, P., Huang, L., and Mennes, M. (2010).
\newblock Fast function-on-scalar regression with penalized basis expansions.
\newblock {\em The International Journal of Biostatistics} {\bf 6,} 28--28.

\bibitem[\protect\citeauthoryear{Resnick and Howard}{Resnick and
  Howard}{2002}]{resnick2002diabetes}
Resnick, H.~E. and Howard, B.~V. (2002).
\newblock Diabetes and cardiovascular disease.
\newblock {\em Annual review of medicine} {\bf 53,} 245--267.

\bibitem[\protect\citeauthoryear{Rodbard}{Rodbard}{2016}]{rodbard2016continuous}
Rodbard, D. (2016).
\newblock Continuous glucose monitoring: a review of successes, challenges, and
  opportunities.
\newblock {\em Diabetes technology \& therapeutics} {\bf 18,} S2--3.

\bibitem[\protect\citeauthoryear{Rooney, Aurora, Wang, Selvin, and
  Punjabi}{Rooney et~al.}{2021}]{rooney2021rationale}
Rooney, M.~R., Aurora, R.~N., Wang, D., Selvin, E., and Punjabi, N.~M. (2021).
\newblock Rationale and design of the hyperglycemic profiles in obstructive
  sleep apnea (hypnos) trial.
\newblock {\em Contemporary Clinical Trials} {\bf 101,} 106248.

\bibitem[\protect\citeauthoryear{Ruppert, Wand, and Carroll}{Ruppert
  et~al.}{2003}]{ruppert2003semiparametric}
Ruppert, D., Wand, M.~P., and Carroll, R.~J. (2003).
\newblock {\em Semiparametric regression}.
\newblock Number~12. Cambridge university press.

\bibitem[\protect\citeauthoryear{Saeedi, Petersohn, Salpea, Malanda, Karuranga,
  Unwin, Colagiuri, Guariguata, Motala, Ogurtsova, et~al\mbox{.}}{Saeedi
  et~al.}{2019}]{saeedi2019global}
Saeedi, P., Petersohn, I., Salpea, P., Malanda, B., Karuranga, S., Unwin, N.,
  Colagiuri, S., Guariguata, L., Motala, A.~A., Ogurtsova, K., et~al. (2019).
\newblock Global and regional diabetes prevalence estimates for 2019 and
  projections for 2030 and 2045: Results from the international diabetes
  federation diabetes atlas.
\newblock {\em Diabetes research and clinical practice} {\bf 157,} 107843.

\bibitem[\protect\citeauthoryear{Scheipl, Staicu, and Greven}{Scheipl
  et~al.}{2015}]{scheipl2015functional}
Scheipl, F., Staicu, A.-M., and Greven, S. (2015).
\newblock Functional additive mixed models.
\newblock {\em Journal of Computational and Graphical Statistics} {\bf 24,}
  477--501.

\bibitem[\protect\citeauthoryear{Sobrin, Green, Sim, Jensen, Tai, Tay, Wang,
  Mitchell, Sandholm, Liu, et~al\mbox{.}}{Sobrin
  et~al.}{2011}]{sobrin2011candidate}
Sobrin, L., Green, T., Sim, X., Jensen, R.~A., Tai, E.~S., Tay, W.~T., Wang,
  J.~J., Mitchell, P., Sandholm, N., Liu, Y., et~al. (2011).
\newblock Candidate gene association study for diabetic retinopathy in persons
  with type 2 diabetes: the candidate gene association resource (care).
\newblock {\em Investigative ophthalmology \& visual science} {\bf 52,}
  7593--7602.

\bibitem[\protect\citeauthoryear{Wood}{Wood}{2011}]{mgcvpackage}
Wood, S.~N. (2011).
\newblock Fast stable restricted maximum likelihood and marginal likelihood
  estimation of semiparametric generalized linear models.
\newblock {\em Journal of the Royal Statistical Society (B)} {\bf 73,} 3--36.

\bibitem[\protect\citeauthoryear{Wood}{Wood}{2017}]{wood2017generalizedbook}
Wood, S.~N. (2017).
\newblock {\em Generalized additive models: an introduction with R}.
\newblock CRC press.

\bibitem[\protect\citeauthoryear{Wood, Goude, and Shaw}{Wood
  et~al.}{2015}]{wood2015generalized}
Wood, S.~N., Goude, Y., and Shaw, S. (2015).
\newblock Generalized additive models for large data sets.
\newblock {\em Journal of the Royal Statistical Society: Series C: Applied
  Statistics} pages 139--155.

\bibitem[\protect\citeauthoryear{Wood, Li, Shaddick, and Augustin}{Wood
  et~al.}{2017}]{wood2017generalized}
Wood, S.~N., Li, Z., Shaddick, G., and Augustin, N.~H. (2017).
\newblock Generalized additive models for gigadata: modeling the uk black smoke
  network daily data.
\newblock {\em Journal of the American Statistical Association} {\bf 112,}
  1199--1210.

\bibitem[\protect\citeauthoryear{Zimmet, Alberti, and Shaw}{Zimmet
  et~al.}{2001}]{zimmet2001global}
Zimmet, P., Alberti, K., and Shaw, J. (2001).
\newblock Global and societal implications of the diabetes epidemic.
\newblock {\em Nature} {\bf 414,} 782--787.

\end{thebibliography}

\end{document}